\documentclass[conference]{IEEEtran}
\usepackage[left=1.52cm,right=1.52cm,top=1.9cm]{geometry}

\usepackage{amsfonts,amsthm}
\usepackage[dvipsnames]{xcolor}
\usepackage{subfigure}

\usepackage{algpseudocode}

\ifCLASSINFOpdf
  \usepackage[pdftex]{graphicx}
\else
\fi
\usepackage{amsmath}
\usepackage{cite}

\newcommand\myatop[2]{\genfrac{}{}{0pt}{}{#1}{#2}} 

\begin{document}
\title{Routing and Spectrum Allocation in\\ Broadband Degenerate EPR-Pair Distribution}
\author{Rohan Bali\textsuperscript{*}, Ashley Tittelbaugh\textsuperscript{*},  Shelbi L. Jenkins\textsuperscript{*},
Anuj Agrawal\textsuperscript{\dag}\\ 
Jerry Horgan\textsuperscript{\ddag}, Marco Ruffini\textsuperscript{\dag}, Daniel Kilper\textsuperscript{\ddag}, Boulat A. Bash\textsuperscript{*}

\\ \textsuperscript{*}Electrical and Computer Engineering Department, University of Arizona, Tucson, AZ, USA
\\ \textsuperscript{\dag}School of Computer Science and Statistics, CONNECT Centre, Trinity College Dublin, Dublin, Ireland
\\ \textsuperscript{\ddag}Electronic and Electrical Engineering Department, CONNECT Centre, Trinity College Dublin, Dublin, Ireland \\
}

\maketitle

\begin{abstract}
We investigate resource allocation for quantum entanglement distribution over an optical network. 
We characterize and model a network architecture that employs a single quasi-deterministic time-frequency heralded EPR-pair source, and develop a routing scheme for distributing entangled photon pairs over such a network.
We focus on fairness in entanglement distribution, and compare both the performance of various spectrum allocation schemes as well as their Jain index.
\end{abstract}

\IEEEpeerreviewmaketitle

\section{Introduction}
Quantum entanglement distribution over a network is essential for large-scale quantum computing, quantum sensing, and quantum security. 
Although various protocols have been proposed (surveyed in \cite{Singh}), the entanglement `source in the middle' approach is a common link configuration for efficient distribution receiving much recent attention.
One example employs a broadband degenerate quasi-deterministic time-frequency heralded Einstein-Podolsky-Rosen (EPR) pair source \cite{Chen}.
In order to form a network from such a source, wavelength selective routing can be used to distribute the broadband entangled photon pairs to consumers at different network nodes. 

The scheme from \cite{Chen} was recently introduced and has the advantage of producing EPR pairs that are heralded in time and frequency, however, it presents unique challenges in routing and spectrum allocation.
The source in \cite{Chen} is degenerate: it outputs entangled photon pairs on the same wavelength.
Thus, photons from a given pair cannot use the same fiber span in the same direction without routing ambiguity or requiring time multiplexing.
Furthermore, although the source is broadband, when segmented into narrow-band channels, the average number of entangled photon pairs it generates per channel varies across the spectrum.
This, along with the path-dependent photon losses, requires the development of novel spectrum allocation strategies.

Fortunately, for the single-source problem considered here, routing and spectrum allocation can be addressed separately.
In this paper we adapt Suurballe's algorithm \cite{Suurballe, Banerjee} to find an optimal route in polynomial time.
We desire fair spectrum allocation, where each node pair receives the same number of EPR pairs on average.
Unfortunately, as in classical optical networks \cite{Chatterjee}, this is an NP-hard integer linear program (ILP).
Therefore, we investigate the performance of various approximation algorithms, and compare them to the optimal ILP solution on a small toy network.
Our analysis also addresses the source placement problem, i.e., finding the optimal location for our entangled photon source. This allows us to analyze both the fairness with which different algorithms can supply heralded EPR pairs throughout the network and the properties of the ideal locations of the source node.

Previous experiments \cite{Wengerowsky} used fully passive wavelength demultiplexing to distribute EPR pairs across a network of four users, showing the viability of such networks. However, the passive fabric limits the ability to redistribute the wavelength spectrum of the source to optimize for consumer demands across a larger network. More recently, this approach has been extended through the use of wavelength selective switching to allow for adaptive allocation of spectrum across a network. By using hyper-entangled states in polarization and frequency along with quantum enabled reconfigurable add-drop multiplexers, \cite{Clark} demonstrated active switching to allocate higher bandwidth entanglement channels to different nodes, and also to demutiplex the channels at the nodes themselves. The degenerate approach \cite{Chen}, however, has not been realized in experiment nor studied from a wavelength routing perspective.  

Section \ref{sec:system_model} overviews the source and network architectures and their models. Section \ref{sec:algorithims} discusses our approaches for optimizing routing and spectrum allocation. Section \ref{sec:results_and_discussion} compares our approaches numerically. We discuss implications of our results and future work in Section \ref{sec:conclusion}.

\section{System Model}
\label{sec:system_model}

\begin{figure}[!t]
\centering
\includegraphics[width=0.8\columnwidth]{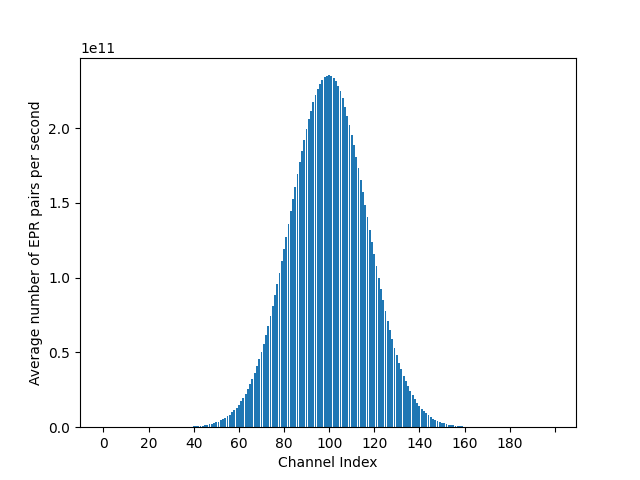}
\caption{Rate of EPR pair generation in 200 channels.}
\label{fig:dist_200}
\end{figure}

\subsection{Broadband Degenerate EPR-pair Generation}
\label{subsec:epr_gen}

We assume the availability of a broadband, quasi-deterministic EPR-pair source. An example of such is the zero-added loss entangled multiplexing (ZALM) scheme described in \cite{Chen}. It employs dual spontaneous parametric down-conversion (SPDC) processes.
This source heralds entangled photon pairs by wavelength demultiplexing the broadband spectrum and detecting coincidence counts occurring at the same wavelength for two idler photons each generated by individual SPDC processes. The corresponding heralded signal photons of now known and identical wavelength are directed through a wavelength division multiplexed (WDM) transmission system with wavelength selective add-drop capability. 

Each SPDC source produces entangled photons at the rate that follows a Gaussian function that we assume is centered at 1550 nm with a full-width half max of 9 nm. As depicted in Fig. \ref{fig:dist_200}, the spectrum is segmented into $m=200$ wavelength channels, each covering a 0.1 nm wavelength range. The central channel is positioned at the peak of the Gaussian distribution. Thus, the lowest and highest indexed channels correspond to a center frequency of 195.9 THz and 191.1 THz with bandwidths of 12.8 GHz and 12.2 GHz respectively. Wavelength-division multiplexing (WDM) prior to Bell measurement enables heralding of the generated EPR pair's WDM channel number. The channels are spaced 0.1 nm apart to prevent fidelity loss from wavelength ambiguity. Due to the Gaussian relationship, the average EPR pair generation rate per second for channels near 1550 nm will be higher than for those on the edges of the spectrum. We include the block diagram of the ZALM source in Fig. \ref{fig:components_source}, but note that our analysis that follows can be adapted to other methods of generating wavelength-heralded EPR pairs.
The output spectrum from this source can be routed and distributed across the network, using wavelength routing techniques like those that have been developed in classical optical transmission networks.

\subsection{Node Architecture}
\label{subsec:node_arch}

\begin{figure}[!t]
\centering
\includegraphics[width=2.5in]{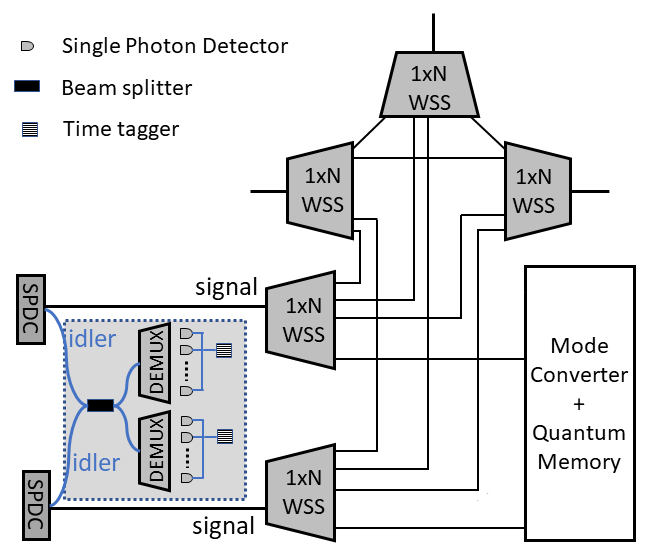}
\caption{Source node.}
\label{fig:components_source}
\end{figure}

\begin{figure}[!t]
\centering
\includegraphics[width=2.5in]{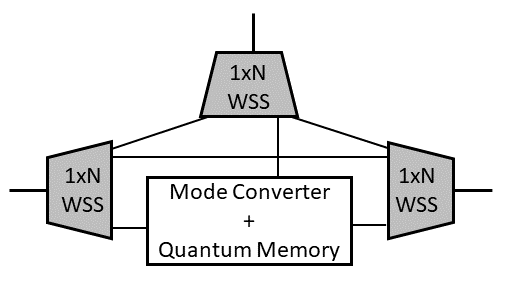}
\caption{Consumer node.}
\label{fig:components_pure_consumer}
\end{figure}

Fig.~\ref{fig:components_source} shows the components of the source node. Each photon of the generated EPR pair is directed into a separate fiber. The node is built around a number of $1\times N$ wavelength selective switches (WSS), whose role is to route wavebands towards different consumer nodes or else towards its own quantum memory bank. These wavebands will route all photons contained within the source wavelength channels depicted in Fig.~\ref{fig:dist_200} that fall within that waveband. Information heralded by the EPR pair generation process (including the channel and timestamp of the generated pair) is transmitted along a classical network that is not depicted here. A consumer node lacks EPR pair generation capability, but has all the other components of the source node, as shown in Fig.~\ref{fig:components_pure_consumer}. An entangled photon arriving at a consumer node can either be directed into its quantum memory or routed to another consumer node. Measured insertion loss $l_{\text{WSS}}$ on Lumentum's TrueFlex Twin WSS ranges from 4 dB to 8 dB \cite{Lumentum}. Hence we analyze EPR pair distribution for two values of WSS loss: $l_{\text{WSS}}\in\{4,8\}$ dB. Loss $l$ in dB is related to power transmittance by $\eta=10^{-l/10}$. 

\subsection{Network Topologies}
\label{subsec:net_topologies}
\begin{figure}[!t]
\centering
\includegraphics[width=2.2in]{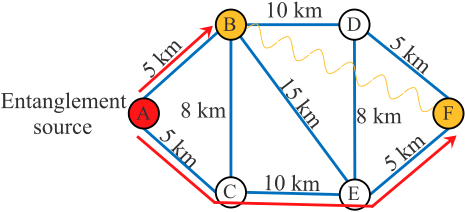}
\caption{Topology of the simple network.}
\label{fig:ecoc_layout}
\end{figure}

Our main topology model is an existing incumbent local exchange carrier (ILEC) node map of Manhattan \cite{Yu, Li}. This topology contains $n=17$ ILEC sites, with each site connected to between 2 and 16 other nodes. The layout of these nodes is shown in Fig.~\ref{fig:manhattan_layout}. While this is the reference topology for validating the performance of our heuristics, the comparison with an optimal ILP solution is restricted to a smaller network topology with $n=6$ nodes, shown in Fig.~\ref{fig:ecoc_layout}, because the optimal fair allocation of EPR pairs is an NP-hard problem.

\subsection{Network Architecture}
\label{subsec:network_arch}

\begin{figure}[!t]
\centering
\includegraphics[width=2.3in]{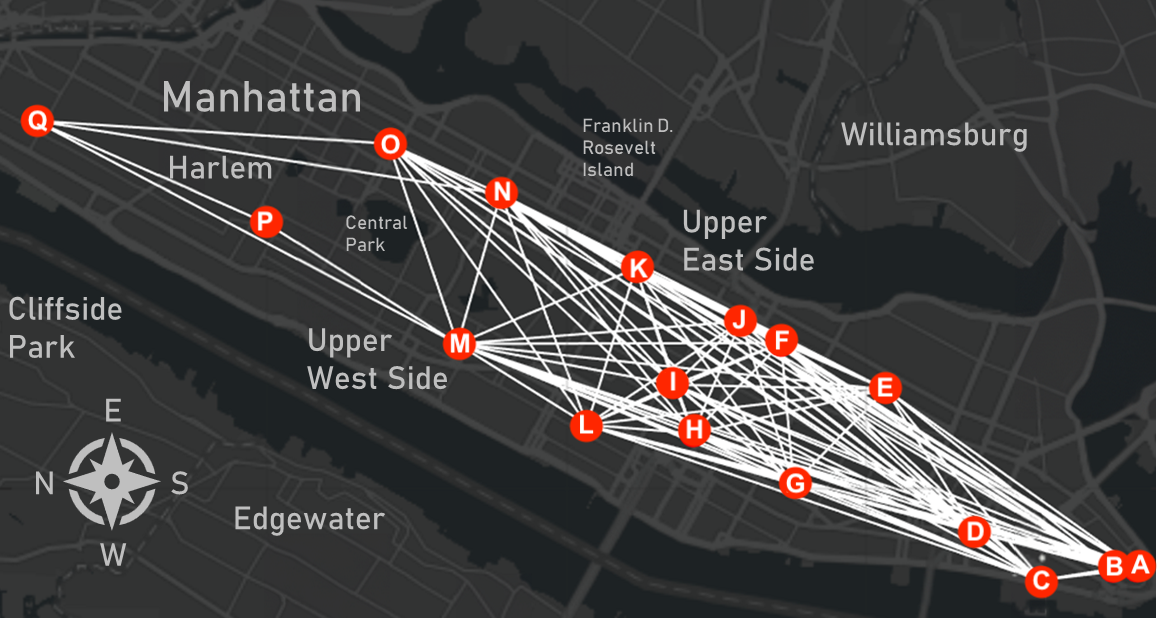}
\caption{A map of Manhattan with ILEC nodes and links overlaid.}
\label{fig:manhattan_layout}
\end{figure}

The deployed fiber link lengths between nodes in the ILEC topology depicted in Fig.~\ref{fig:manhattan_layout} are unknown.  Thus, we use direct ``as the crow flies'' distance as a proxy. Standard single mode fiber is assumed on each link. We employ a higher loss coefficient of $\alpha=0.4$ dB/km than typical fiber loss at 1550 nm (found in, e.g.,\cite{Kingfisher}) to account for higher losses and longer run lengths characteristic of metro fiber plant. We assume that all pairs of nodes in the network request EPR pairs from the source. Each wavelength channel is assigned to a single pair. The wavelength routing mechanism follows a circuit-switching approach. The routes serving different sets of node node pairs do not interfere with one another, however, photons from a particular channel cannot be directed to two different nodes of a pair via the same fiber in the same direction, as this results in a routing ambiguity.
Therefore, we only consider networks that allow disjoint light-paths from the source to each of the $k$ pairs of consumer nodes.

\subsection{Network Model}
\label{subsec:network_model}

We represent a network as a graph denoted by $\mathcal{G}=(\mathcal{V},\mathcal{E})$, where $\mathcal{V}$ and $\mathcal{E}$ are the sets of vertices and directed edges, respectively.  
We also define a map $w:\mathcal{E}\to\mathbb{R}$ that assigns photon losses (in dB) as edge weights. 
We construct $\mathcal{G}$ for the network topologies described in Section \ref{subsec:net_topologies} as follows:
\begin{itemize}
\item For each pair $(i,j)$ of connected consumer nodes we add the following directed edges and the corresponding vertices: $e_{i,j}\equiv\left(v_{i,j}^{\text{(out)}}, v_{j,i}^{\text{(in)}}\right)$ and $ e_{j,i}\equiv \left(v_{j,i}^{\text{(out)}}, v_{i,j}^{\text{(in)}}\right)$ to $\mathcal{G}$. Superscripts indicate whether the vertices describe incoming or outgoing ports. The weight of the edges is $w\left(e_{i,j}\right)=w\left(e_{j,i}\right)=\alpha\times d(i,j)$, where $d(i,j)$ is the distance (in km) between nodes $i$ and $j$, and $\alpha$ is optical fiber loss (in dB/km) discussed in Section \ref{subsec:network_arch}.
\item For each consumer node $i$, we iterate over all nodes $j,k$ that connect to $i$, and add edges $e_{i,j,k}\equiv \left(v_{i,j}^{\text{(in)}}, v_{i,k}^{\text{(out)}}\right)$ to $\mathcal{E}$. This captures the consumer nodes' internal connections between incoming and outgoing ports.  Since the photons routed through a consumer node must traverse two WSSes, the weight of these edges is $w\left(e_{i,j,k}\right)=2l_{\rm WSS}$, as discussed in Section \ref{subsec:network_arch}.  Furthermore, we also add edges $e_{i,i,j}\equiv \left(v_{i,j}^{\text{(in)}}, v_{i}^{\text{(mem)}}\right)$ describing internal connections to node $i$’s quantum memory to $\mathcal{E}$, and the corresponding vertices to $\mathcal{V}$. Since only one WSS is traversed in this case, $w\left(e_{i,i,j}\right)=l_{\rm WSS}$.
\item For the source node $s$, we iterate over all nodes $j$ that connect to $s$, and add edges $e_{s,j}\equiv \left(v_{s,j}^{\text{(out)}}, v_{j, s}^{\text{(in)}}\right)$ to $\mathcal{E}$ and corresponding vertices to $\mathcal{V}$. The weight of these edges is $w\left(e_{s,j}\right)=\alpha\times d(s,j)$. Consumer nodes incoming vertices $v_{j,s}^{\text{(in)}}$ are connected to outgoing vertices and quantum memories as described above. Finally, we add edges $e_{j}\equiv \left(v_{s}^{(gen)}, v_{s,j}^{\text{(out)}}\right)$ and $e_{s}\equiv \left(v_{s}^{(gen)}, v_{s}^{\text{(mem)}}\right)$ from vertex $v_{s}^{\text{(gen)}}$ describing EPR pair generator to all outgoing ports and vertex $v_{s}^{\text{(mem)}}$ describing source's own quantum memory. The weights for these edges are $w\left(e_{j}\right)=2l_{\rm WSS}$ and $w\left(e_{s}\right)=l_{\rm WSS}$, per above. Note that the source node does not have incoming ports.
\end{itemize}
The total loss on a path from source to a consumer node $i$ is the sum of weights of the edges connecting $v_{s}^{\text{(gen)}}$ to $v_i^{\text{(mem)}}$.

\subsection{Max-min (Egalitarian) Fairness}
\label{subsec:egal_fairness}
We seek max-min, or egalitarian, fairness, and maximize the minimum average rate of EPR pairs received by all $k = n(n-1)/2$ pairs $(i,j)$ of $n$ nodes \cite{Bezakova}.  Let $l_{(i,j)}$ be the total loss (in dB) from the source to nodes $(i,j)$. That is, $l_{(i,j)}$ is the sum of losses on the disjoint paths from source to nodes $i$ and $j$, per Section \ref{subsec:network_model}. Then, transmittance $\eta_{(i,j)}=10^{-l_{(i,j)}/10}$ is the fraction of the entangled photon pairs that are received by $(i,j)$. Let $\mathcal{A}_{(i,j)}$ be the set of channels assigned to node pair $(i,j)$. Since each channel cannot be assigned to more than one node pair, the set $\mathcal{P}=\left\{\mathcal{A}_{(i,j)}:i,j=1,\ldots,n, i \neq j \right\}$ partitions the $m$ available channels. Let $\bar{n}_{x}$ be the average rate of EPR pair generated in channel $x$. The average rate of EPR pairs received by node pair $(i,j)$ is then $\bar{n}_{(i,j)}=\eta_{(i, j)}\sum_{x\in \mathcal{A}_{(i,j)}} \bar{n}_{x}$ and the max-min fair allocation involves the following optimization: 
   $\max_{\mathcal{P}} \min_{(i, j)} \bar{n}_{(i,j)}$.

\section{Algorithms}
\label{sec:algorithims}

Orthogonality of sets $\mathcal{A}_{(i,j)}$ allows treating routing and spectrum allocation problems separately, as discussed next. 

\subsection{Optimal routing}
\label{sec:paths}

Unlike standard networks, our `source-in-the-middle' entanglement distribution system described in Section \ref{sec:system_model} requires two disjoint light paths from source $s$ to nodes $i$ and $j$ that minimize total loss $l_{(i,j)}$ for each pair $(i,j)$ in the network. Per Section \ref{subsec:network_model}, this translates to finding edge-disjoint routes in $\mathcal{G}$ from $v_s^{\text{(gen)}}$ to $v_i^{\text{(mem)}}$ and  $v_j^{\text{(mem)}}$ minimizing sum of weights of these paths. 
To this end, we use Suurballe’s algorithm \cite{Suurballe, Banerjee} as follows: for each pair $(i,j)$ we add a dummy vertex $v_{i,j}^{\text{(d)}}$ to $\mathcal{V}$ and dummy zero weighted edges: $e_{i,j}^{\text{(d)}}\equiv \left(v_i^{\text{(mem)}}, v_{i,j}^{\text{(d)}}\right)$ and $e_{j,i}^{\text{(d)}}\equiv \left(v_j^{\text{(mem)}}, v_{i,j}^{\text{(d)}}\right)$ to $\mathcal{E}$. Suurballe's algorithm yields two edge-disjoint paths of minimum total weight between $v_s^{\text{(gen)}}$ and $v_{i,j}^{\text{(d)}}$.
Removing dummy vertices and edges returns edge-disjoint paths of minimum total weight from $v_{s}^{\text{(gen)}}$ to $v_{i}^{\text{(mem)}}$ and $v_{j}^{\text{(mem)}}$ for all pairs $(i,j)$.
Suurballe's algorithm is polynomial-time in graph size.

\subsection{Spectrum Allocation Strategies}

Let $X$ be an $m\times n(n-1)/2$ binary matrix with $X_{x,(i,j)}=1$ if channel $x$ is assigned to node pair $(i,j)$ and zero otherwise (note that the pair $(i,j)$ indexes columns of $X$). Formally, $X_{x,(i,j)}=\left\{1~\text{if}~x\in\mathcal{A}_{(i,j)}; 0~\text{else}\right\}$.
Also define an $n(n-1)/2\times n(n-1)/2$ diagonal matrix $\Lambda$ with transmittances $\eta_{(i,j)}^\ast$ of optimal routes (see Section \ref{sec:paths}) from source to each $(i,j)$ on the diagonal and a vector $N = \left[\bar{n}_{1}, \ldots, \bar{n}_{m}\right]$ of average EPR-pair-generation rates (see Section \ref{subsec:egal_fairness}).
For some $X$, the average rate of EPR pairs received by $(i,j)$ is $\bar{n}_{(i,j)}=\left[NX\Lambda\right]_{(i,j)}$, the $(i,j)^{\text{th}}$ entry of vector $NX\Lambda$.

Finding an optimal spectrum allocation matrix $X$ is a well-known problem in optical networking \cite{Chatterjee}. Here we focus on maintaining max-min fairness in `source-in-the-middle' entanglement distribution.

\subsubsection{Optimal Assignment}

The following integer linear program (ILP) yields the optimal max-min fair solution $T_{\text opt}$:
\begin{IEEEeqnarray}{ll}
\label{eq:ilp}
\IEEEyesnumber  \IEEEyessubnumber*
\max_X T \text{~s.t.~}&\sum_{\myatop{i,j=1}{i\neq j}}^{n} X_{x,(i,j)} = 1, \forall x=1,\ldots,m \label{eq:ilp_second}\\
&\left[NX\Lambda\right]_{(i,j)} \geq T, \forall i,j=1,\ldots,n,i\neq j,  \label{eq:ilp_third}
\end{IEEEeqnarray}
where constraint \eqref{eq:ilp_second} enforces that each channel is assigned only once and \eqref{eq:ilp_third} ensures that each node pair receives EPR pair rate of at least $T$. 
Since ILP is NP-hard, we consider various approximations.

\subsubsection{First Fit \cite{Chatterjee}}
We assign channels sequentially to a node pair. If EPR pair rate $T$ is reached, then we repeat for the next node pair. We restart with a smaller $T$ if channels are exhausted before all node pairs attain EPR pair rate $T$. 

\subsubsection{Round Robin}
The channels are assigned one at a time to node pairs in the descending order of generated ERP pair rate \cite{Aziz}. 

\subsubsection{Random}
Each request is assigned roughly the same number of channels at random.

\subsubsection{Modified Longest Processing Time First (LPT) \cite{Graham, Deuermeyer}}

This is a well-known machine scheduling algorithm. We modify it to greedily optimize for the max-min rather than min-max goal: each channel is assigned to a node pair which maximizes the current minimum average received EPR pair rate across the node pairs.
While our experiments indicate that this substantially improves this algorithm in our setting, we have not derived any analytical performance guarantees.

\subsubsection{$1/(m-k+1)$-approximation \cite{Bezakova}}

This iterative polynomial-time algorithm converges to a solution that is guaranteed to be within $1/(m-k+1)$ of the optimal max-min value, where, in our setting $m$ is the number of channels and $k=n(n-1)/2$ is the number of node pairs.
We make two modifications: 1) instead of always assigning one channel to each node pair in each round, we allow skipping a channel assignment; 2) in each round, we prefer the assignment which minimizes the total rate of EPR pair generation that is assigned.
These are invoked as long as it does not impact the overall max-min value, hence they can only increase the minimum received EPR pair rate for all node pairs, all the while preserving the original guarantee of $1/(m-k+1)$.

\subsubsection{$\max(0, T_{\text{f-opt}} - \max_{(i,j),x}\eta_{(i,j)}\bar{n}_{x})$-assignment guarantee \cite{Bezakova}} 
The minimum average EPR pair rate received by a node pair guaranteed by this algorithm is limited by the maximum EPR pair rate any node pair can receive:  $\max(0, T_{\text{f-opt}} - \max_{(i,j),x}\eta_{(i,j)}\bar{n}_{x})$, where $T_{\text{f-opt}}$ is the optimal solution to the integer linear program in \eqref{eq:ilp} relaxed to allow fractional channel assignments.
This algorithm first solves a linear program to obtain a fractional channel assignment, and then resolves  assignments to multiple requests. 

\begin{figure*}[!th]
  \centering
    \subfigure{\raisebox{1\height}{Legend for Figs.~\ref{fig:simple} and \ref{fig:manhattan}:} \includegraphics[width=1.5\columnwidth]{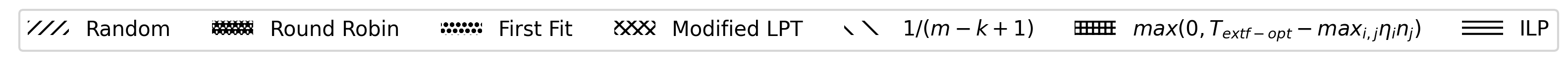}}
  \addtocounter{subfigure}{-1}
  \subfigure[Minimum average received EPR pair rates]{\includegraphics[scale=0.42]{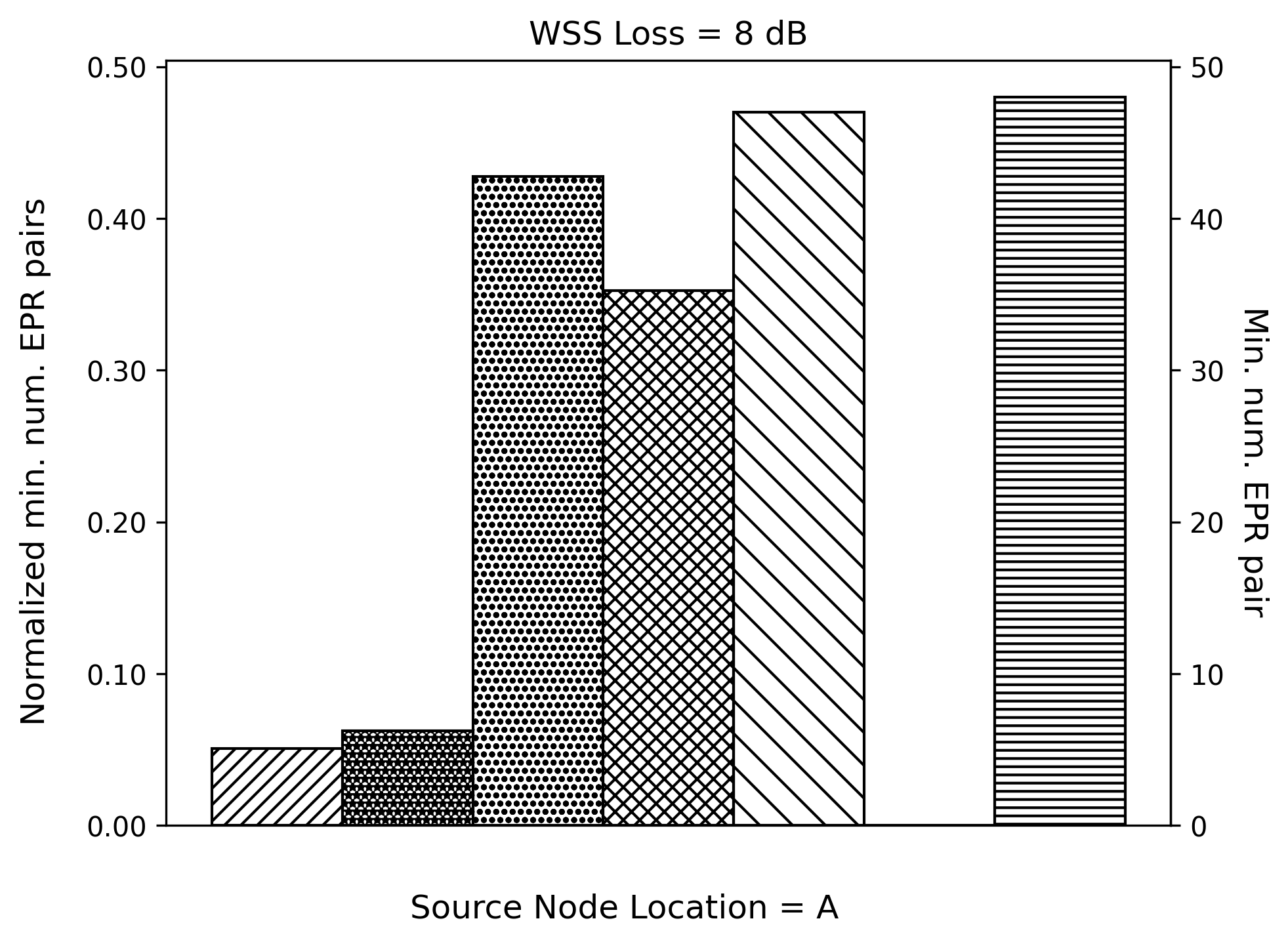} \label{fig:simple_results}}
  \subfigure[Jain Index]{\includegraphics[scale=0.42]{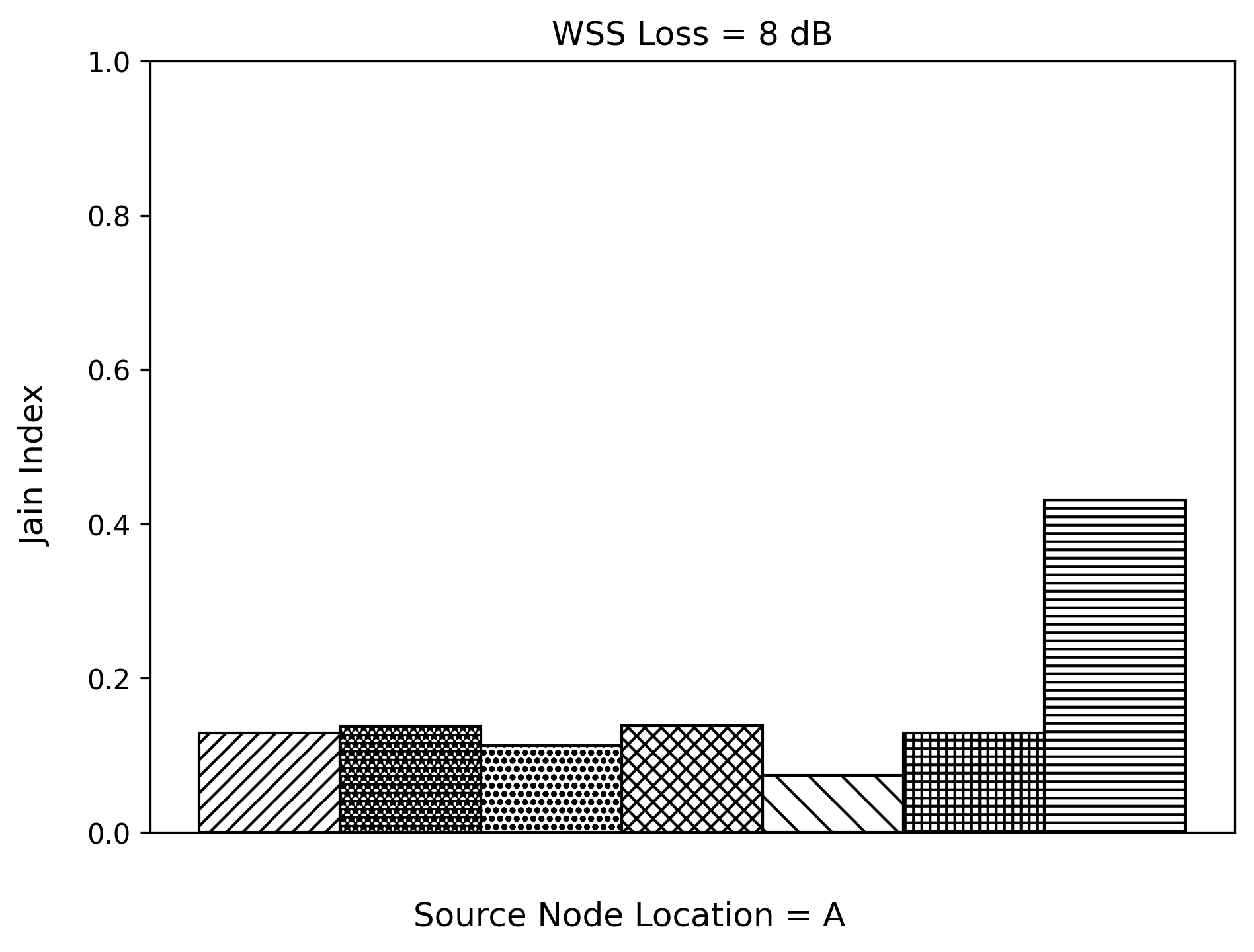} \label{simple_jain}}
  \caption{Comparison of performance using different allocation strategies on the simple network depicted in Fig.~\ref{fig:ecoc_layout} for 8 dB WSS loss. \ref{fig:simple_results} shows normalized (left-axis label) and unnormalized (right-axis label) minimum average EPR pair rates received by any node pair. \label{fig:simple}}
\end{figure*}

\begin{figure*}[!th]
  \centering
  \subfigure[Minimum average received EPR pair rates]{\includegraphics[width=0.94\columnwidth]{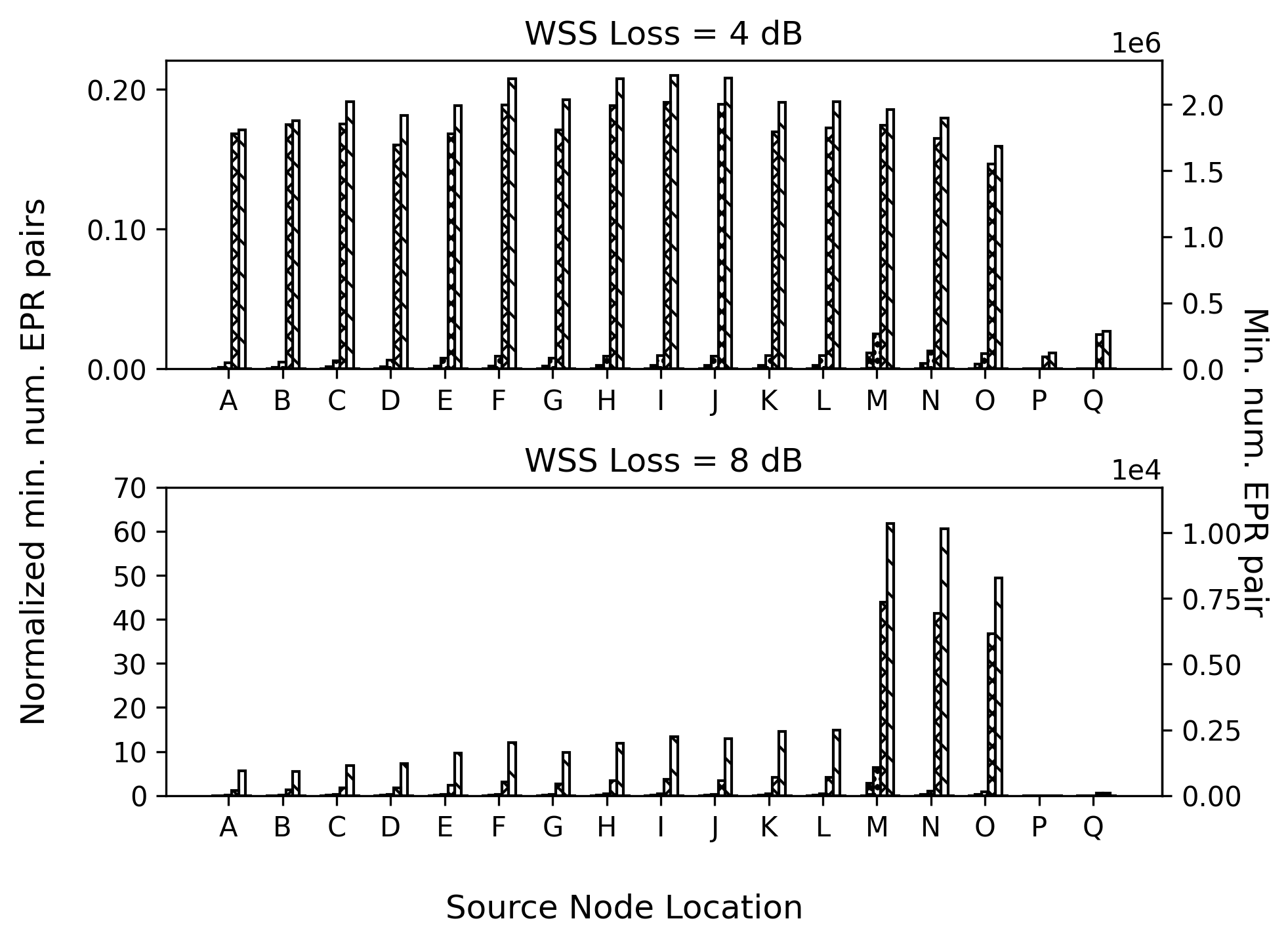}\label{fig:manhattan_results}}
  \hfill
  \subfigure[Jain index]{\includegraphics[width=0.94\columnwidth]{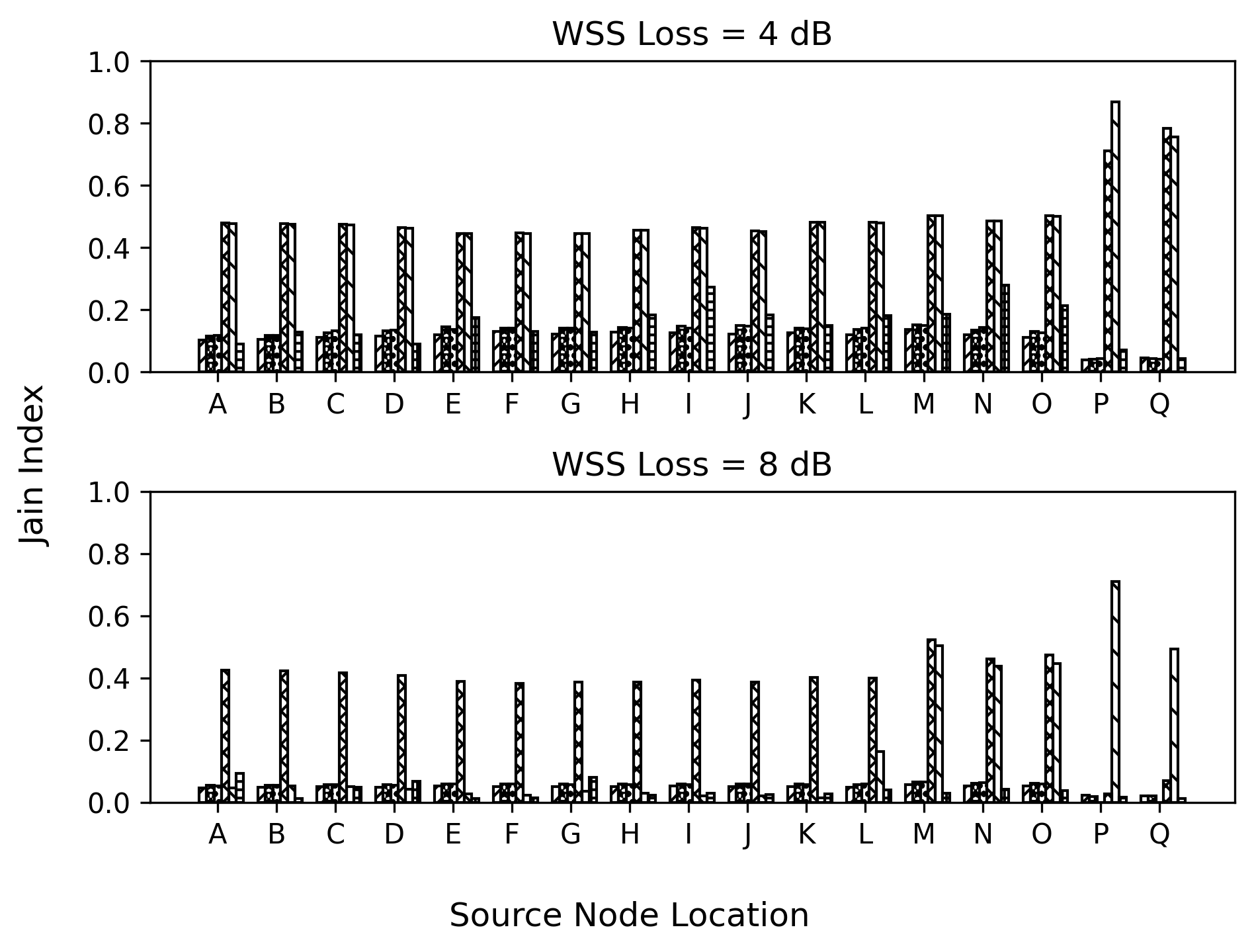} \label{fig:manhattan_jain}}
  \caption{Comparison of performance using different allocation strategies on the ILEC network depicted in Fig.~\ref{fig:manhattan_layout} for 4 dB and 8 dB WSS loss. \ref{fig:manhattan_results} shows normalized (left-axis label) and unnormalized (right-axis label) minimum average EPR pair rates received by any node pair.\label{fig:manhattan}}
\end{figure*}

\section{Results and Discussion}
\label{sec:results_and_discussion}

In Figs.~\ref{fig:simple_results} and \ref{fig:manhattan_results} we report unnormalized and normalized minimum average received EPR pair rates for topologies described in Section \ref{subsec:net_topologies}. Normalization is with respect to the highest-loss consumer pair's photon count across source node locations, when assigned all channels, i.e.: $\min_{(i,j)}\eta_{(i,j)}\sum_{x=1}^{m}\bar{n}_{x}$.
In Figs.~~\ref{simple_jain} and \ref{fig:manhattan_jain} we report the Jain index \cite{Jain} $\frac{\sum_{(i, j)} \bar{n}_{(i,j)})^2}{\frac{n(n-1)}{2}\sum_{(i, j)} \bar{n}_{(i,j)}^{2}}$, which ranges from 1 (completely fair) to $\frac{1}{n(n-1)/2}$ (most unfair).

The first-fit and round-robin algorithms are sensitive to the node pair order. The random assignment can yield varying performance metrics across different executions. Also, although the ILP algorithm consistently produces the same minimum number of assigned average EPR pair rates in each run, it may use distinct assignment configurations, resulting in varying Jain index. Thus, the results for the first-fit, random, round-robin, and ILP algorithms are averaged over 1000 runs, with each run randomizing the order of processing the node pairs. The error bars are negligibly small and are not depicted.

Fig.~\ref{fig:simple_results} depicts the minimum average EPR pair rates received by any node pair in the simple network topology depicted in Fig.~\ref{fig:ecoc_layout} when placing the source at node A and a WSS loss of 8 dB. 
We can calculate the optimal solution using ILP for this topology. We note that the $1/(m-k+1)$ approximation algorithm is close to optimal. Modified LPT and First-Fit algorithms perform well; the First-Fit algorithm's performance is surprising given its relative simplicity. Random and round-robin algorithms perform poorly. The $\max(0, T_{\text{f-opt}} - \max_{(i,j),x}\eta_{(i,j)}\bar{n}_{x})$ algorithm shows the poorest performance on this metric since it does not assign any channels to some node pairs. 

Fig.~\ref{simple_jain} shows the performance of these strategies on the Jain index. The ILP solution, which optimizes for the minimum average received EPR pair rate, also performs the best on this fairness measure. The performance of the other strategies is comparable to each other. Interestingly, the $1/(m-k+1)$ approximation algorithm which performed well for minimum photons received, performs the poorest on the Jain index. This underscores a constraint of the Jain index as a metric, as it solely evaluates the relative fairness among assignments without taking into account the quantity of EPR pairs allocated.

Fig.~\ref{fig:manhattan_results} shows the normalized and unnormalized minimum average EPR pair rates that any node pair can receive in the ILEC network topology depicted in Fig.~\ref{fig:manhattan_layout} when the source is positioned at different locations in the network. These results are for two distinct cases, with WSS losses of 4 dB and 8 dB. Due to the complexity of the ILP program for this topology, we cannot calculate the optimal solution here.

The number of intermediate nodes traversed by a path in the ILEC network varies significantly based on the source node location. A linear increase in the number of intermediate nodes traversed leads to an exponential shift in the associated transmittance of the path. Hence we see that minimum average EPR pair rates vary significantly across source node locations. Also due to this exponential relationship between path loss in dB and transmittance, the difference in the minimum average EPR pair rates across source node locations is accentuated when the WSS loss is set to 8dB as opposed to 4 dB. 

Normalized values for the minimum number of EPR pairs facilitate comparisons across diverse network topologies, as in the case of comparison between results in \ref{fig:simple_results} and results in \ref{fig:manhattan_results} with an 8 dB loss. However, the reference for normalization is affected when the WSS loss is changed. Decreasing the WSS loss from 8 dB to 4 dB significantly changes the scale of values after normalization, as can be seen in the two sub-graphs in \ref{fig:manhattan_results}. This behavior arises because, while the ratio between path losses stays roughly constant when WSS losses are doubled, normalization here captures the ratio between the transmittance of the paths, which has an exponential relationship with dB loss. As a result highly connected graphs perform better at higher losses, as seen in the figure. 

Consistent with the findings displayed for the basic network, both the $1/(m-k+1)$ algorithm and modified LPT algorithms demonstrate effectiveness in optimizing the minimum average ERP pair rates received by a node pairs. However, First-Fit algorithm, which performed well on the simple topology from Fig.~\ref{fig:ecoc_layout}, performs poorly here. This is due to the greater disparity in the `value' of a channel to different node pairs in the ILEC network, owing to the greater difference in path losses to these node pairs. For the First-Fit algorithm, situations may arise where no 'high value' channels are available by the time a node pair with highly lossy paths reaches its turn.

Fig.~\ref{fig:manhattan_jain} shows the performance of these strategies on the Jain index. 
The $1/(m-k+1)$ algorithm and LPT strategy both show the best performance on this metric. Interestingly, despite not assigning any channels to some requests, $\max(0, T_{\text{f-opt}} - \max_{(i,j),x}\eta_{(i,j)}\bar{n}_{x})$ strategy performs better on the Jain index relative to many others. The $1/(m-k+1)$ strategy exhibits superior performance compared to other approaches on source nodes $P$ and $Q$ when the WSS loss is at 8 dB.

Across different source locations, we see those with higher nodal degrees can supply higher minimum average EPR rates throughout the network. Nodes $A$ through $L$ have degree 14, and show a similar performance. Node $M$ has the highest degree (16) and shows the best performance. Nodes $P$ and $Q$ have degree two and four, respectively, and demonstrate the poorest performance. Interestingly, the performance of nodes $N$ and $O$ which have degree 15 show dramatic improvement over nodes with degree 14. This can be attributed to the fact that these node's neighbours are neighbors to node $Q$ and second-order neighbors to node $P$; two nodes that have few other neighbors. Thus while the nodes with 14 neighbors cannot efficiently supply EPR pairs when one of the nodes is $P$ or $Q$, source nodes $N$ and $O$ do not suffer from this problem.

\section{Conclusion}
\label{sec:conclusion}
In this study, we explore the optimization of EPR pair distribution in quantum networks to address the increasing demand for efficient quantum computation and communication. We consider a novel source-in-the-middle time-frequency heralded EPR source and examine multiple allocation algorithms for fair routing and spectrum distribution while considering factors like loss reduction and resource management. Findings reveal that different allocation strategies have varying impacts on minimum average EPR pair rate allocations and fairness. In simple and complex network topologies, the $1/(m-k+1)$ approximation and modified LPT algorithms prove effective. The Jain index underscores the trade-offs between optimal solutions and fairness metrics, emphasizing the need for careful consideration in quantum network deployment, resource distribution, and real-world applications in quantum computing, sensing, and secure communication. Future work should focus on algorithm refinement and experimental implementations.

\section*{Acknowledgment}

This material is based upon work supported by the National Science Foundation under Grant No. CNS-2107265 and Science Foundation Ireland grants 20/US/3708, 21/US-C2C/3750, and 13/RC/2077\_P2.

\end{document}